\documentstyle[aps,epsf,graphicx,eqsecnum,preprint]{revtex}
\newcommand{\be}{\begin{equation}}
\newcommand{\ee}{\end{equation}}
\newcommand{\bea}{\begin{eqnarray}}
\newcommand{\eea}{\end{eqnarray}}
\def \in{\leftskip = 40 pt\rightskip = 40pt}

\def \out{\leftskip = 0 pt\rightskip = 0pt}
\def\nn{\nonumber\\}

\def\frak#1#2{{\textstyle{{#1}\over{#2}}}}

\def\semi{;\hfil\break}

\def\GeV{{\rm GeV}}
\def\TeV{{\rm TeV}}
\def\Ycal{{\cal Y}}
\def\Atbar{{\overline{A}_t}}
\def\Abbar{{\overline{A}_b}}
\def\Ataubar{{\overline{A}_{\tau}}}
\def\Abar{{\overline{A}}}

\def\Cbar{{\overline{C}}}
\def\Ybar{{\overline{\Ycal}}}
\def\Qbar{{\overline{Q}}}

\def\DRED{\ifmmode{{\rm DRED}} \else{{DRED}} \fi}
\def\DREDp{\ifmmode{{\rm DRED}'} \else{${\rm DRED}'$} \fi}  
\def\NSVZ{\ifmmode{{\rm NSVZ}} \else{{NSVZ}} \fi}

\def\npb{{Nucl.\ Phys.\ }{\bf B}}
\def\prd{{Phys.\ Rev.\ }{\bf D}}

\def\plb{{Phys.\ Lett.\ }{\bf B}}

\def\thbar{\bar\theta}

\def\bxhat{\hat\beta_{\xi}}
\def\lf{16\pi^2}
\def\llf{(16\pi^2)^2}
\def\lllf{(16\pi^2)^3}
\def\Tr{\hbox{Tr}}

\def\ttil{\tilde t}
\def\btil{\tilde b}
\def\tautil{\tilde \tau} 
\def\util{\tilde u}
\def\dtil{\tilde d}
\def\etil{\tilde e}

\def\nutil{\tilde \nu}
\def\chitil{\tilde \chi}
\begin{document}

\begin{titlepage}
\begin{flushright}
LTH 489\\
hep-ph/0010301\\
\end{flushright}

\vspace*{3mm}

\begin{center}
{\Huge
The Fayet-Iliopoulos $D$-term and its renormalisation in the MSSM}\\[12mm]
 
{\bf I.~Jack and D.R.T.~Jones} \\
   
 
\vspace{8mm} 

Dept. of Mathematical Sciences,
University of Liverpool, Liverpool L69 3BX, UK\\
\end{center}

\vspace{3mm}
\begin{abstract}

We consider the renormalisation of the  Fayet-Iliopoulos $D$-term in a
softly-broken  supersymmetric gauge theory with a non-simple gauge group
containing an abelian factor, and present the
associated $\beta$-function through three loops.  We also include in an 
appendix the result for
several abelian factors. We specialise to the 
case of the minimal supersymmetric standard model (MSSM), and investigate the
behaviour of the Fayet-Iliopoulos coupling $\xi$ for various boundary conditions
at the unification scale. We focus particularly on the case
of non-standard soft supersymmetry breaking couplings, for which $\xi$
evolves significantly between the unification scale and the weak scale.

\end{abstract}

\end{titlepage}

\section{Introduction}

In abelian gauge theories with $N=1$ supersymmetry 
there exists a possible 
invariant that is not allowed in the non-abelian case: the 
Fayet-Iliopoulos $D$-term,
\be
L =
\xi\int V (x,\theta, \thbar)\,d^4\theta  = \xi D(x).
\label{dta}
\ee
In previous papers \cite{xione}\cite{xitwo} we have discussed the 
renormalisation of $\xi$ in the presence of 
the standard soft supersymmetry-breaking terms 
\be
L_{\rm SB}=-(m^2)^j_i\phi^{i}\phi_j-
\left(\frac{1}{6}h^{ijk}\phi_i\phi_j\phi_k+\frac{1}{2}b^{ij}\phi_i\phi_j
+ \frac{1}{2}M\lambda\lambda+{\rm h.c.}\right)
\label{Aaf}
\ee
The result for $\beta_{\xi}$ is as follows:
\be
\beta_{\xi} = \frac{\beta_g}{g}\xi + \bxhat
\label{exacta}
\ee
where $\bxhat$ is determined by $V$-tadpole (or in components 
$D$-tadpole) graphs, and is independent of $\xi$. 
Although in Refs.~\cite{xione}, \cite{xitwo} we restricted ourselves to the 
abelian case, 
it is evident that a $D$-term can occur with a direct product gauge group
($G_1\otimes G_2\cdots$) if there is an  abelian factor: 
as is the case for the  MSSM. In the MSSM context one may 
treat $\xi$ as a free parameter at the weak scale\cite{gouv}, 
in which case there is no need to know $\bxhat$. However, if 
we know $\xi$ at gauge unification, for example, 
then we need $\bxhat$ to predict 
$\xi$ at low energies. Our purpose in this paper is first of all to give
the result for $\bxhat$ through three loops for a general direct product gauge 
group. For simplicity of exposition, we restrict ourselves in the main body of
the paper to the case of one abelian factor, postponing the more general 
result (which is complicated by the possibility of ``kinetic mixing''\cite{kin}
between
different abelian factors) to an Appendix. 
We shall then specialise to the case of the MSSM, and perform some 
running analyses to determine the size of $\xi(M_Z)$ for various choices of 
boundary conditions at the unification scale $M_X$.

\section{The General Case}
First of all, for completeness and to establish notation,  let us recapitulate 
the standard results for the supersymmetric theory. 
We take an $N=1$ supersymmetric gauge theory with gauge group 
$\Pi_{\alpha} G_{\alpha}$
and with superpotential
\be
W(\Phi)=\frac{1}{6}Y^{ijk}\Phi_i\Phi_j\Phi_k+
\frac{1}{2}\mu^{ij}\Phi_i\Phi_j.
\label{eqf}
\ee
We will be assuming here that the gauge group has one abelian factor, which we 
shall 
take to be $G_1$. We shall denote the hypercharge matrix for $G_1$ by $\Ycal$.
At one loop we have
\begin{mathletters}
\bea
\label{beone}
\lf\beta_{g_{\alpha}}^{(1)} &=& g_{\alpha}^3Q_{\alpha} =
g_{\alpha}^3\left[T(R_{\alpha})-3C(G_{\alpha})\right], \label{beone:1}\\
\lf\gamma^{(1)i}{}_j &=& P^i{}_j
=\frac{1}{2}Y^{ikl}Y_{jkl}-2\sum_{\alpha}g_{\alpha}^2[C(R_{\alpha})]^i{}_j,
\label{beone:2}
\eea
\end{mathletters}
where $R_{\alpha}$ is the group representation for $G_{\alpha}$ acting on the
chiral fields,  $C(R_{\alpha})$ the corresponding quadratic Casimir and 
$T(R_{\alpha}) = (r_{\alpha})^{-1}\Tr [C(R_{\alpha})]$ , 
$r_{\alpha}$ being  the dimension of $G_{\alpha}$. For the adjoint 
representation, $C(R_{\alpha})=C(G_{\alpha})I_{\alpha}$, where $I_{\alpha}$ is
the $r_{\alpha}\times r_{\alpha}$ unit matrix. 
Note that $T(R_1)=\Tr[\Ycal^2]$, $\left[C(R_1)\right]^i{}_j=(\Ycal^2)^i{}_j$.
At two loops we have 
\begin{mathletters}
\bea
\label{betwo}
\llf\beta_{g_{\alpha}}^{(2)}&=&2g_{\alpha}^5C(G_{\alpha})Q_{\alpha}
-2g_{\alpha}^3r_{\alpha}^{-1}\Tr\left[P C(R_{\alpha})\right]
\label{betwo:1}\\
\llf\gamma^{(2)i}{}_j&=&[-Y_{jmn}Y^{mpi}-2\sum_{\alpha}g_{\alpha}^2
C(R_{\alpha})^p{}_j\delta^i{}_n]P^n{}_p+
2\sum_{\alpha}g_{\alpha}^4C(R_{\alpha})^i{}_jQ_{\alpha}.
\label{betwo:2}
\eea
\end{mathletters}
For completeness and later reference, we also quote here the general result
for $\beta_{g_{\alpha}}^{\DRED(3)}$, which is a straightforward generalisation 
of the result of Ref.~\cite{jjn}:
\bea
\beta_{g_{\alpha}}^{\DRED(3)}&=&
3r_{\alpha}^{-1}g_{\alpha}^3Y^{ikm}Y_{jkn}P^n{}_mC(R_{\alpha})^j{}_i
+6r_{\alpha}^{-1}g_{\alpha}^3\sum_{\beta}g_{\beta}^2
\Tr\left[PC(R_{\alpha})C(R_{\beta})\right]\nn
&+&r_{\alpha}^{-1}g_{\alpha}^3\Tr\left[P^2C(R_{\alpha})\right]
-6r_{\alpha}^{-1}g^3_{\alpha}\sum_{\beta}Q_{\beta}g_{\beta}^4
\Tr\left[C(R_{\alpha})C(R_{\beta})\right] \nn
&-& 4r_{\alpha}^{-1}g_{\alpha}^5C(G_{\alpha})\Tr\left[PC(R_{\alpha})\right]
+g_{\alpha}^7Q_{\alpha}C(G_{\alpha})\left[4C(G_{\alpha})-Q_{\alpha}\right],
\label{bethree}
\eea
 We recall that 
gauge anomaly cancellation
requires 
\be\Tr[\Ycal C(R_{\alpha})]=0\label{anom}
\ee
and naturalness (or cancellation of $U_1$-gravitational anomalies)
requires 
\be
\Tr[\Ycal]= 0. \label{gravnat}
\ee

The diagrams contributing to $\bxhat$ through three loops for a general 
non-simple gauge group
are essentially the same as those depicted for the pure abelian case   
in Ref.~\cite{xitwo}, but reinterpreting internal gauge and gaugino 
propagators as ranging over all gauge groups in the direct product.
Potential new 3-loop graphs 
(involving a 3-point gauge vertex, or a gauge/gaugino
vertex) give contributions which vanish due to anomaly cancellation
(such as $C(G_{\alpha})\Tr[\Ycal C(R_{\alpha})]$). It is then relatively
easy to generalise the abelian result to the general case. 
We find 
\be
\lf\bxhat^{(1)} = 2g_1\Tr\left[\Ycal m^2\right] \label{exacto}
\ee
\be
\lf\bxhat^{(2)}=-4g_1\Tr\left[\Ycal m^2 \gamma^{(1)}\right],\label{exactb}
\ee
\bea
\lllf\frac{\bxhat^{(3)\DREDp}}{g_1} &=& -6(\lf)^2\Tr\left[\Ycal m^2
\gamma^{(2)}\right]-4\Tr\left[\Ycal WP\right]- 
\frac{5}{2}\Tr\left[\Ycal HH^{\dagger}\right]\nn
&+& 2\Tr\left[\Ycal P^2m^2\right]
-24\zeta(3)\sum_{\alpha}g_{\alpha}^2\Tr\left[\Ycal WC(R_{\alpha})\right]\nn
&+& 12\zeta(3)\sum_{\alpha}g_{\alpha}^2\Tr\left[\Ycal M_{\alpha}^*H 
C(R_{\alpha}) 
+ \hbox{c.c.}\right]\nn
&-&96\zeta(3)\sum_{\alpha,\beta}g_{\alpha}^2g_{\beta}^2M_{\alpha}M^*_{\alpha}
\Tr[\Ycal C(R_{\alpha})C(R_{\beta})]\nn
&-&24\zeta(3)\left\{\sum_{\alpha,\beta}g_{\alpha}^2g_{\beta}^2M_{\alpha}
M^*_{\beta}\Tr[\Ycal C(R_{\alpha})C(R_{\beta})]+\hbox{c.c.}\right\}\nn
\label{exactc}
\eea
where\cite{jj} 
\be
W^i{}_j =
(\frac{1}{2}Y^2m^2 +\frac{1}{2}m^2Y^2 +h^2)^i{}_j
+2Y^{ipq}Y_{jpr}(m^2)^r{}_q -8\sum_{\beta}g_{\beta}^2M_{\beta}
M_{\beta}^*C(R_{\beta})^i{}_j,\label{Wdef}
\ee

\be
H^i{}_j=h^{ikl}Y_{jkl}+4\sum_{\beta}g_{\beta}^2M_{\beta}[C(R_{\beta})]^i{}_j
\label{Hdef}
\ee
with
$(Y^2)^i{}_j =  Y^{ikl}Y_{jkl}, (h^2)^i{}_j =  h^{ikl}h_{jkl}.$
These results are computed using the \DREDp scheme, which is a variant
of DRED defined so as to ensure that $\beta$-functions for physical
couplings have no  dependence on the $\epsilon$-scalar mass\cite{jjmvy}.
Most of the  terms in Eq.~(\ref{exactc}) correspond in a simple way to
the  analogous terms in Eq.~(5.2) of Ref.~\cite{xitwo}; the only subtle
point  being the $MM^*g^4$ terms, where one sees easily that only in the
case  of Fig.~15(e) of \cite{xitwo} can the two gaugino masses belong to
different  gauge groups $(G_{\alpha})$. Thus the last term in
Eq.~(\ref{exactc}),  and the $MM^*g^4$ terms from the terms involving $H$, 
come entirely from this  particular figure. 

It was proved in Ref.~\cite{xione} 
in the pure abelian case that if the 
$m^2$ dependence in $\bxhat$ takes the form $\Tr[m^2A]$, then we have
\be
\Tr[\Ycal A]=2{\beta_{g_1}\over g_1^2}.\label{gxi}
\ee
It is easy to see that the proof extends to the direct product case, and indeed
we can check Eq.~(\ref{gxi}) explicitly using 
Eqs.~(\ref{exacto})--(\ref{exactc}) and (\ref{beone:1}), (\ref{betwo:1}) and 
(\ref{bethree}).

\section{The MSSM}
We now specialise to the case of the MSSM.  
The relevant part of the MSSM superpotential is:
\be
W = H_2 t^c Y_t Q  + H_1 b^c Y_b Q + H_1 \tau^c Y_{\tau} L
\ee
where $Y_t$, $Y_b$, $Y_{\tau}$ are $3\times 3$ Yukawa flavour matrices.

The gauge $\beta$-functions are given at one loop by
\be
\lf\beta_{g_{\alpha}}=b_{\alpha}g_{\alpha}^3,
\ee
where 
\be
b_1=\frak{33}{5},\quad b_2=1,\quad b_3=-3,\ee 
and our $U_1$ coupling normalisation corresponds to the usual one such that 
$g_1^2 = \frac{5}{3}(g')^2$.
For the anomalous dimensions of the chiral superfields we have
at one loop:

\bea
\lf\gamma^{(1)}_{t^c} &=& P_{t^c}=2Y_tY_t^{\dagger} -2C_{t^c},\nn
\lf\gamma^{(1)}_{b^c} &=& P_{b^c}=2Y_bY_b^{\dagger} -2C_{b^c},\nn
\lf\gamma^{(1)}_{Q} &=& P_Q=Y_b^{\dagger}Y_b + Y_t^{\dagger}Y_t -2C_Q,\nn
\lf\gamma^{(1)}_{\tau^c} &=& P_{\tau^c}
=2Y_{\tau}Y_{\tau}^{\dagger}-2C_{\tau^c},\nn
\lf\gamma^{(1)}_{L} &=& P_L=Y_{\tau}^{\dagger}Y_{\tau} -2C_L,
\nn
\lf\gamma^{(1)}_{H_1} &=& P_{H_1}=\Tr \left[Y_{\tau}^{\dagger}Y_{\tau}
+3Y_b^{\dagger}Y_b\right]-2C_L
,\nn
\lf\gamma^{(1)}_{H_2} &=& P_{H_2}=3\Tr \left[Y_t^{\dagger}Y_t\right]-2C_L,
\label{Pone}
\eea
where
\bea
C_{t^c}&=&\frak{4}{3}g_3^2 +\frak{4}{15}g_1^2,\nn
C_{b^c}&=&\frak{4}{3}g_3^2 +\frak{1}{15}g_1^2,\nn
C_Q&=&\frak{4}{3}g_3^2 + \frak{3}{4}g_2^2+\frak{1}{60}g_1^2,\nn
C_{\tau^c}&=&\frak{3}{5}g_1^2,\nn
C_L&=&\frak{3}{4}g_2^2 +\frak{3}{20}g_1^2.
\eea
At two loops\cite{bjork} the anomalous dimensions are given by:
\begin{mathletters}
\bea
\label{gatwo}
\llf\gamma_{t^c}^{(2)}&=&-2Y_t\left(P_Q+P_{H_2}\right)Y_t^{\dagger}
-2P_{t^c}C_{t^c}+2(\frak{4}{15}b_1g_1^4+\frak{4}{3}b_3g_3^4),\label{gatwo:a}\\
\llf\gamma_{b^c}^{(2)}&=&-2Y_b\left(P_Q+P_{H_1}\right)Y_b^{\dagger}
-2P_{b^c}C_{b^c}+2(\frak{1}{15}b_1g_1^4+\frak{4}{3}b_3g_3^4),\label{gatwo:b}\\
\llf\gamma_Q^{(2)}&=&-Y_t^{\dagger}\left(P_{t^c}+P_{H_2}\right)Y_t
-Y_b^{\dagger}\left(P_{b^c}+P_{H_1}\right)Y_b\nn
&-&2P_QC_Q+2(\frak{1}{60}b_1g_1^4+ \frak{3}{4}b_2g_2^4 + \frak{4}{3}b_3g_3^4),
\label{gatwo:c}\\
\llf\gamma_{\tau^c}^{(2)}&=&-2Y_{\tau}\left(P_L+P_{H_1}\right)Y_{\tau}^{\dagger}
-2P_{\tau^c}C_{\tau^c}+\frak{6}{5}b_1g_1^4,\label{gatwo:d}\\
\llf\gamma_L^{(2)}&=&-Y_{\tau}^{\dagger}[P_{\tau^c}+P_{H_1}]Y_{\tau}
-2P_LC_L+\frak{3}{10}b_1g_1^4 + \frak{3}{2}b_2g_2^4,\label{gatwo:e}\\
\llf\gamma_{H_1}^{(2)}&=&-3\Tr[Y_bP_QY_b^{\dagger}+Y_b^{\dagger}P_{b^c}Y_b]
-\Tr[Y_{\tau}P_QY_{\tau}^{\dagger}+Y_{\tau}^{\dagger}P_{\tau^c}Y_{\tau}]\nn
&-&2C_LP_{H_1}+\frak{3}{10}b_1g_1^4 + \frak{3}{2}b_2g_2^4,\label{gatwo:f}\\
\llf\gamma_{H_2}^{(2)}&=&-3\Tr[Y_tP_QY_t^{\dagger}+Y_t^{\dagger}P_{t^c}Y_t]
-2C_LP_{H_2}+\frak{3}{10}b_1g_1^4 + \frak{3}{2}b_2g_2^4.\label{gatwo:g}
\eea
\end{mathletters}
We now turn to the soft couplings. The quantities $W$ and $H$ defined in 
Eqs.~(\ref{Wdef}), (\ref{Hdef}) are given by
\bea
W_{t^c}&=& (2m_{t^c}^2 + 4 m_{H_2}^2 ) Y_t Y_t^{\dagger}
+ 4Y_t m_Q^2Y_t^{\dagger}
+ 2Y_t Y_t^{\dagger}  m_{t^c}^2
+ 4 h_t h_t^{\dagger}-8C_{t^c}^{MM},\nn
W_{b^c}&=& (2m_{b^c}^2 + 4 m_{H_1}^2 ) Y_b Y_b^{\dagger}
+ 4Y_b m_Q^2Y_b^{\dagger}
+ 2Y_b Y_b^{\dagger}  m_{b^c}^2
+ 4 h_b h_b^{\dagger}-8C_{b^c}^{MM},\nn
W_Q&=& (m_Q^2 + 2 m_{H_2}^2 ) Y_t^{\dagger} Y_t
+ (m_Q^2 + 2 m_{H_1}^2 ) Y_b^{\dagger} Y_b
+ [ Y_t^{\dagger} Y_t + Y_b^{\dagger} Y_b ] m_Q^2
+ 2 Y_t^{\dagger} m_{t^c}^2 Y_t\nn
&+& 2 Y_b^{\dagger} m_{b^c}^2 Y_b
+ 2 h_t^{\dagger} h_t + 2 h_b^{\dagger} h_b-8C_Q^{MM},\nn
W_{\tau^c}&=& (2m_{\tau^c}^2 + 4 m_{H_1}^2 ) Y_{\tau} Y_{\tau}^{\dagger}
+ 4Y_{\tau} m_L^2Y_{\tau}^{\dagger}
+ 2Y_{\tau} Y_{\tau}^{\dagger}  m_{\tau^c}^2
+ 4 h_{\tau} h_{\tau}^{\dagger}-8C_{\tau^c}^{MM},\nn
W_L&=& (m_L^2 + 2 m_{H_1}^2 ) Y_{\tau}^{\dagger} Y_{\tau}
+ 2 Y_{\tau}^{\dagger} m_{\tau^c}^2 Y_{\tau}
+ Y_{\tau}^{\dagger} Y_{\tau}  m_L^2
+ 2 h_{\tau}^{\dagger} h_{\tau}-8C_L^{MM},\nn
W_{H_1}&=& {\rm Tr} \Bigl [6(m_{H_1}^2 + m_Q^2) Y_b^{\dagger} Y_b
   + 6m_{b^c}^2  Y_bY_b^{\dagger} 
+2(m_{H_1}^2 + m_L^2)Y_{\tau}^{\dagger} Y_{\tau}
+ 2 Y_{\tau}^{\dagger} m_{\tau^c}^2 Y_{\tau}\nn
&+& 6h_b^{\dagger}h_b
+2h_{\tau}^{\dagger}
h_{\tau}\Bigr]-8C_L^{MM},\nn
W_{H_2}&=&6 {\rm Tr} [(m_{H_2}^2 + m_Q^2)Y_t^{\dagger} Y_t
+ m_{t^c}^2 Y_tY_t^{\dagger} 
+ h_t^{\dagger} h_t ]-8C_L^{MM},\label{Wone}
\eea
where
\bea
C^{MM}_{t^c}&=&\frak{4}{3}|M_3|^2g_3^2 +\frak{4}{15}|M_1|^2g_1^2,\nn   
C^{MM}_{b^c}&=&\frak{4}{3}|M_3|^2g_3^2 +\frak{1}{15}|M_1|^2g_1^2,\nn
C^{MM}_Q&=&\frak{4}{3}|M_3|^2g_3^2 + 
\frak{3}{4}|M_2|^2g_2^2+\frak{1}{60}|M_1|^2g_1^2,\nn
C^{MM}_{\tau^c}&=&\frak{3}{5}|M_1|^2g_1^2,\nn
C^{MM}_L&=&\frak{3}{4}|M_2|^2g_2^2 +\frak{3}{20}|M_1|^2g_1^2,
\eea
and
\bea
H_{t^c}&=&4h_tY_t^{\dagger}+4C_{t^c}^M,\nn
H_{b^c}&=&4h_bY_b^{\dagger}+4C_{b^c}^M,\nn
H_Q&=&2(Y_t^{\dagger}h_t+Y_b^{\dagger}h_b)+4C_Q^M,\nn
H_{\tau^c}&=&4h_{\tau}Y_{\tau}^{\dagger}+4C_{\tau^c}^M,\nn
H_L&=&2Y_{\tau}^{\dagger}h_{\tau}+4C_L^M,\nn
H_{H_1}&=&\Tr[6Y_b^{\dagger}h_b+Y_{\tau}^{\dagger}h_{\tau}]+4C_L^M,\nn
H_{H_2}&=&6\Tr[Y_t^{\dagger}h_t]+4C_L^M,
\label{Hone}
\eea
where
\bea
C^M_{t^c}&=&\frak{4}{3}M_3g_3^2 +\frak{4}{15}M_1g_1^2,\nn
C^M_{b^c}&=&\frak{4}{3}M_3g_3^2 +\frak{1}{15}M_1g_1^2,\nn
C^M_Q&=&\frak{4}{3}M_3g_3^2 + \frak{3}{4}M_2g_2^2+\frak{1}{60}M_1g_1^2,\nn  
C^M_{\tau^c}&=&\frak{3}{5}M_1g_1^2,\nn
C^M_L&=&\frak{3}{4}M_2g_2^2 +\frak{3}{20}M_1g_1^2.
\eea
With all these subsidiary definitions we can now give the results for 
$\bxhat$ up to three loops. We have
\be
\lf\bxhat^{(1)} =2\sqrt{\frak{3}{5}}g_1
\Tr[m_Q^2-m_L^2 -2m_{t^c}^2 +m_{b^c}^2 +m_{\tau^c}^2 -m_{H_1}^2 + m_{H_2}^2]
\label{bxione}
\ee
\bea
\llf\bxhat^{(2)}&=&-4\sqrt{\frak{3}{5}}g_1\Tr(m^2_QP_Q -m^2_LP_L
-2m^2_{t^c}P_{t^c}+m^2_{b^c}P_{b^c}\nn
&+&m^2_{\tau^c}P_{\tau^c}-m^2_{H_1}P_{H_1}+m^2_{H_2}P_{H_2})\label{bxitwo}
\eea
\bea
\lllf\bxhat^{(3)}&=&\sqrt{\frak{3}{5}}g_1\Bigl[-6\llf\beta^{(3)}_{\xi_1}
-4\beta^{(3)}_{\xi_2}
-\frak{5}{2}\beta^{(3)}_{\xi_3}+2\beta^{(3)}_{\xi_4}\nn
&+&\zeta(3)(-24\beta^{(3)}_{\xi_5}
+12\beta^{(3)}_{\xi_6}-96\beta^{(3)}_{\xi_7}-48\beta^{(3)}_{\xi_8})\Bigr],
\eea
where
\bea
\beta^{(3)}_{\xi_1}&=&\Tr(m^2_Q\gamma_Q^{(2)} -m^2_L\gamma_L^{(2)}
-2m^2_{t^c}\gamma_{t^c}^{(2)}+m^2_{b^c}\gamma_{b^c}^{(2)}+m^2_{\tau^c}\gamma_{\tau^c}^{(2)}
-m^2_{H_1}\gamma_{H_1}^{(2)} +m^2_{H_2}\gamma_{H_2}^{(2)}),\nn
\beta^{(3)}_{\xi_2}&=&\Tr(W_QP_Q -W_LP_L
-2W_{t^c}P_{t^c}+W_{b^c}P_{b^c}+W_{\tau^c}P_{\tau^c}-W_{H_1}P_{H_1}+W_{H_2}P_{H_2}),\nn
\beta^{(3)}_{\xi_3}&=&\Tr(H_Q^{\dagger}H_Q -H_L^{\dagger}H_L
-2H_{t^c}^{\dagger}H_{t^c}+H_{b^c}^{\dagger}H_{b^c}+H_{\tau^c}^{\dagger}H_{\tau^c}
-H_{H_1}^{\dagger}H_{H_1}+H_{H_2}^{\dagger}H_{H_2}),\nn
\beta^{(3)}_{\xi_4}&=&\Tr(m^2_QP_Q^2 -m^2_LP_L^2-2m^2_{t^c}P_{t^c}^2
+m^2_{b^c}P_{b^c}^2 +m^2_{\tau^c}P_{\tau^c}^2 
-m^2_{H_1}P_{H_1}^2 +m^2_{H_2}P_{H_2}^2),\nn
\beta^{(3)}_{\xi_5}&=&\Tr(W_QC_Q -W_LC_L
-2W_{t^c}C_{t^c}+W_{b^c}C_{b^c}+W_{\tau^c}C_{\tau^c}-W_{H_1}C_L+W_{H_2}C_L),\nn
\beta^{(3)}_{\xi_6}&=&\Tr(H_QC^{M*}_Q -H_LC^{M*}_L
-2H_{t^c}C^{M*}_{t^c}+H_{b^c}C^{M*}_{b^c}+H_{\tau^c}C^{M*}_{\tau^c}-H_{H_1}C^{M*}_L
+H_{H_2}C^{M*}_L)
+\hbox{c.c.},\nn
\beta^{(3)}_{\xi_7}&=&3(C^{MM}_QC_Q-C^{MM}_LC_L-2C^{MM}_{t^c}C_{t^c}
+C^{MM}_{b^c}C_{b^c}+C^{MM}_{\tau^c}C_{\tau^c})\nn
\beta^{(3)}_{\xi_8}&=&3(|C^M_Q|^2
-|C^M_L|^2-2|C^M_{t^c}|^2
+|C^M_{b^c}|^2+|C^M_{\tau^c}|^2).
\label{xidef}
\eea
We shall now present our MSSM results specialised to the commonly considered 
case where only the 3rd generation Yukawa couplings are significant. We also
take the gaugino masses to be real.
Writing $\lambda_t$, $\lambda_b$ and $\lambda_{\tau}$ for the 3rd generation 
couplings, Eq.~(\ref{Pone}) becomes
\bea
P_{t^c}&=&2\lambda^2_t-2C_{t^c}\nn
P_{b^c} &=&2\lambda^2_b-2C_{b^c}\nn
P_Q&=&\lambda^2_b+\lambda^2_t -2C_Q\nn
P_{\tau^c}&=&2\lambda^2_{\tau}-2C_{\tau^c}\nn
P_L&=&\lambda^2_{\tau}-2C_L\nn
P_{u^c}&=&-2C_{t^c}\nn
P_{d^c} &=&-2C_{b^c}\nn
P_R&=&-2C_Q\nn
P_{e^c}&=&-2C_{\tau^c}\nn
P_N&=&-2C_L\nn
P_{H_1}&=&\lambda^2_{\tau}+3\lambda^2_b-2C_L\nn
P_{H_2}&=&3\lambda^2_t-2C_L,
\eea
where $\{t,b,Q,\tau,L\}$ now refers to the 3rd generation, and $\{u,d,R,e,N\}$
refers to either of the 1st or 2nd generations. Eq.~(\ref{gatwo}) now takes 
the form
\begin{mathletters}
\bea
\label{gatwoa}
\llf\gamma_{t^c}^{(2)}&=&-2\lambda_t^2\left(P_Q+P_{H_2}\right)
-2P_{t^c}C_{t^c}+2(\frak{4}{15}b_1g_1^4+\frak{4}{3}b_3g_3^4),\label{gatwoa:a}\\
\llf\gamma_{b^c}^{(2)}&=&-2\lambda_b^2\left(P_Q+P_{H_1}\right)
-2P_{b^c}C_{b^c}+2(\frak{1}{15}b_1g_1^4+\frak{4}{3}b_3g_3^4),\label{gatwoa:b}\\
\llf\gamma_Q^{(2)}&=&-\lambda_t^2\left(P_{t^c}+P_{H_2}\right)
-\lambda^2_b\left(P_{b^c}+P_{H_1}\right)\nn
&-&2P_QC_Q+2(\frak{1}{60}b_1g_1^4+ \frak{3}{4}b_2g_2^4 + \frak{4}{3}b_3g_3^4),
\label{gatwoa:c}\\
\llf\gamma_{\tau^c}^{(2)}&=&-2\lambda_{\tau}^2\left(P_L+P_{H_1}\right)
-2P_{\tau^c}C_{\tau^c}+\frak{6}{5}b_1g_1^4,
\label{gatwoa:d}\\
\llf\gamma_L^{(2)}&=&-\lambda_{\tau}^2[P_{\tau^c}+P_{H_1}]
-2P_LC_L+\frak{3}{10}b_1g_1^4 + \frak{3}{2}b_2g_2^4,
\label{gatwoa:e}\\
\llf\gamma_{u^c}^{(2)}&=&
-2P_{u^c}C_{t^c}+2(\frak{4}{15}b_1g_1^4+\frak{4}{3}b_3g_3^4),
\label{gatwoa:f}\\
\llf\gamma_{d^c}^{(2)}&=&
-2P_{d^c}C_{b^c}+2(\frak{1}{15}b_1g_1^4+\frak{4}{3}b_3g_3^4),
\label{gatwoa:g}\\
\llf\gamma_R^{(2)}&=&-2P_RC_Q+2(\frak{1}{60}b_1g_1^4+ \frak{3}{4}b_2g_2^4 
+ \frak{4}{3}b_3g_3^4),
\label{gatwoa:h}\\
\llf\gamma_{e^c}^{(2)}&=&
-2P_{e^c}C_{\tau^c}+\frak{6}{5}b_1g_1^4.
\label{gatwoa:i}\\
\llf\gamma_N^{(2)}&=&   
-2P_NC_L+\frak{3}{10}b_1g_1^4 + \frak{3}{2}b_2g_2^4
\label{gatwoa:j}\\
\llf\gamma_{H_1}^{(2)}&=&-3\lambda_{b}^2[P_Q+P_{b^c}]
-\lambda_{\tau}^2[P_L+P_{\tau^c}]\nn
&-&2C_LP_{H_1}+\frak{3}{10}b_1g_1^4 + \frak{3}{2}b_2g_2^4,
\label{gatwoa:k}\\  
\llf\gamma_{H_2}^{(2)}&=&-3\lambda_{t}^2[P_Q+P_{t^c}]
-2C_LP_{H_2}+\frak{3}{10}b_1g_1^4 + \frak{3}{2}b_2g_2^4,
\label{gatwoa:l}
\eea
\end{mathletters}
Correspondingly, we retain only the three 3rd generation trilinear soft
couplings $h_t=A_t\lambda_t$, $h_t=A_b\lambda_b$ and 
$h_{\tau}=A_{\tau}\lambda_{\tau}$. Eq.(\ref{Wone}) now becomes
\bea
W_{t^c}&=& 4\lambda_t^2(m_{t^c}^2 + m_Q^2+ m_{H_2}^2 +A_t^2) -8C_{t^c}^{MM}\nn
W_{b^c}&=& 4\lambda_b^2(m_{b^c}^2 + m_Q^2+m_{H_1}^2 +A_b^2) -8C_{b^c}^{MM}\nn
W_Q&=& 2\lambda_t^2(m_{t^c}^2+m_Q^2 + m_{H_2}^2 +A_t^2) 
+ 2\lambda_b^2(m_{b^c}^2+m_Q^2 + m_{H_1}^2 +A_b^2) -8C_Q^{MM}\nn
W_{\tau^c}&=& 4\lambda_{\tau}^2(m^2_{\tau^c}
+m^2_{H_1}+m^2_L+A_{\tau}^2)-8C_{\tau^c}^{MM}\nn
W_L&=& 2\lambda_{\tau}^2(m^2_{H_1}+m^2_{\tau^c}+m^2_L+A_{\tau}^2)-8C_L^{MM},\nn
W_{u^c}&=& -8C_{t^c}^{MM},\nn
W_{d^c}&=& -8C_{b^c}^{MM},\nn
W_R&=& -8C_Q^{MM},\nn
W_{e^c}&=& -8C_{\tau^c}^{MM},\nn
W_N&=& -8C_L^{MM},\nn
W_{H_1}&=& 6\lambda_b^2(m^2_{H_1}+m^2_{b^c}+m^2_Q+A_b^2)
+2\lambda_{\tau}^2(m^2_{H_1}+m^2_L+m^2_{\tau^c}+A_{\tau}^2)-8C_L^{MM},\nn
W_{H_2}&=&6\lambda_t^2(m^2_{H_2}+m^2_{t^c}+m^2_Q+A_t^2)-8C_L^{MM}.
\eea
Eq.~(\ref{Hone}) now becomes
\bea
H_{t^c}&=&4A_t\lambda_t^2+4C_{t^c}^M,\nn
H_{b^c}&=&4A_b\lambda_b^2+4C_{b^c}^M,\nn
H_Q&=&2(A_t\lambda_t^2+A_b\lambda_b^2)+4C_Q^M,\nn
H_{\tau^c}&=&4A_{\tau}\lambda_{\tau}^2+4C_{\tau^c}^M,\nn
H_L&=&2A_{\tau}\lambda_{\tau}^2+4C_L^M\nn
H_{u^c}&=&4C_{t^c}^M,\nn
H_{d^c}&=&4C_{b^c}^M,\nn
H_R&=&4C_Q^M,\nn
H_{e^c}&=&4C_{\tau^c}^M,\nn
H_N&=&4C_L^M,\nn
H_{H_1}&=&6A_{b}\lambda_b^2+2A_{\tau}\lambda_{\tau}^2+4C_L^M,\nn
H_{H_2}&=&6A_t\lambda_t^2+4C_L^M.
\eea
Eqs.~(\ref{bxione}), (\ref{bxitwo}) now become
\bea 
\lf\bxhat^{(1)} &=&2\sqrt{\frak{3}{5}}g_1
\Bigl(m_Q^2+2m_R^2-m_L^2 -2m_N^2-2m_{t^c}^2 -4m_{u^c}^2+m_{b^c}^2 
+2m_{d^c}^2 \nn
&+&m_{\tau^c}^2 +2m_{e^c}^2-m_{H_1}^2 + m_{H_2}^2\Bigr),
\eea
\bea
\llf\bxhat^{(2)}&=&-4\sqrt{\frak{3}{5}}g_1\Bigl(m^2_QP_Q +2m^2_RP_R-m^2_LP_L
-2m^2_NP_N-2m^2_{t^c}P_{t^c}-4m^2_{u^c}P_{u^c}\nn
&+&m^2_{b^c}P_{b^c}+2m^2_{d^c}P_{d^c}
+m^2_{\tau^c}P_{\tau^c}+2m^2_{e^c}P_{e^c}-m^2_{H_1}P_{H_1}
+m^2_{H_2}P_{H_2}\Bigr).
\eea
Finally, Eq.~(\ref{xidef}) is replaced by
\bea
\beta^{(3)}_{\xi_1}&=&m^2_Q\gamma_Q^{(2)} 
+2m^2_R\gamma_R^{(2)} -m^2_L\gamma_L^{(2)}
-2m^2_N\gamma_N^{(2)} -2m^2_{t^c}\gamma_{t^c}^{(2)}
-4m^2_{u^c}\gamma_{u^c}^{(2)} 
+m^2_{b^c}\gamma_{b^c}^{(2)} +2m^2_{d^c}\gamma_{d^c}^{(2)} \nn
&+&m^2_{\tau^c}\gamma_{\tau^c}^{(2)}
+2m^2_{e^c}\gamma_{e^c}^{(2)} -m^2_{H_1}\gamma_{H_1}^{(2)} 
+m^2_{{H_2}}\gamma_{{H_2}}^{(2)}\nn
\beta^{(3)}_{\xi_2}&=&W_QP_Q +2W_RP_R-W_LP_L-2W_NP_N
-2W_{t^c}P_{t^c}-4W_{u^c}P_{u^c}+W_{b^c}P_{b^c}+2W_{d^c}P_{d^c}\nn
&+&W_{\tau^c}P_{\tau^c}+2W_{e^c}P_{e^c}
-W_{H_1}P_{H_1}+W_{{H_2}}P_{{H_2}}\nn
\beta^{(3)}_{\xi_3}&=&H_Q^2 +2H_R^2-H_L^2-2H_N^2
-2H_{t^c}^2-4H_{u^c}^2+H_{b^c}^2+2H_{d^c}^2\nn
&+&H_{\tau^c}^2+2H_{e^c}^2
-H_{H_1}^2+H_{{H_2}}^2\nn
\beta^{(3)}_{\xi_4}&=&m^2_QP_Q^2 +2m^2_RP_R^2 -m^2_LP_L^2
-2m^2_NP_N^2 -2m^2_{t^c}P_{t^c}^2
-4m^2_{u^c}P_{u^c}^2 +m^2_{b^c}P_{b^c}^2 +2m^2_{d^c}P_{d^c}^2 \nn
&+&m^2_{\tau^c}P_{\tau^c}^2 +2m^2_{e^c}P_{e^c}^2
-m^2_{H_1}P_{H_1}^2 +m^2_{{H_2}}P_{{H_2}}^2\nn
\beta^{(3)}_{\xi_5}&=&W_QC_Q +2W_RC_Q-W_LC_L-2W_NC_L
-2W_{t^c}C_{t^c}-4W_{u^c}C_{t^c}+W_{b^c}C_{b^c}+2W_{d^c}C_{b^c}\nn
&+&W_{\tau^c}C_{\tau^c}+2W_{e^c}C_{\tau^c}
-W_{H_1}C_L+W_{{H_2}}C_L\nn
\beta^{(3)}_{\xi_6}&=&2(H_QC^M_Q +2H_RC^M_Q-H_LC^M_L-2H_NC^M_L
-2H_{t^c}C^M_{t^c}-4H_{u^c}C^M_{t^c}+H_{b^c}C^M_{b^c}+2H_{d^c}C^M_{b^c}\nn
&+&H_{\tau^c}C^M_{\tau^c}+2H_{e^c}C^M_{\tau^c}
-H_{H_1}C^M_L+H_{{H_2}}C^M_L)\nn
\beta^{(3)}_{\xi_7}&=&3(C^{MM}_QC_Q-C^{MM}_LC_L-2C^{MM}_{t^c}C_{t^c}
+C^{MM}_{b^c}C_{b^c}+C^{MM}_{\tau^c}C_{\tau^c})\nn
\beta^{(3)}_{\xi_8}&=&3\left[|C^M_Q|^2
-|C^M_L|^2-2|C^M_{t^c}|^2
+|C^M_{b^c}|^2+|C^M_{\tau^c}|^2\right].
\eea
\section{Running analysis}
As we mentioned in the Introduction, if we have no prejudice as to the value of
$\xi$ at the gauge unification scale $M_X$, then we may as well treat $\xi$ as 
a free parameter
at the weak scale\cite{gouv}, and the running of $\xi$ is irrelevant. However
it is conceivable that the underlying theory at scales beyond $M_X$ may favour
certain values of $\xi(M_X)$, and then the running of $\xi$ would need to be
considered. We shall see that for currently popular choices of boundary
conditions at $M_X$--namely, the minimal supergravity scenario, and
the Anomaly Mediated Supersymmetry Breakdown (AMSB) 
scenario--the running of $\xi$ is determined 
predominantly by the first term on the RHS of Eq.~(\ref{exacta}) 
between 
$M_X$ and $M_Z$, and hence to a good approximation we have
\be
\xi(M_Z)\approx{g_1(M_Z)\over{g_1(M_X)}}\xi(M_X). \label{xiapp}
\ee
For instance, we find from Eqs.~(\ref{exacto}), (\ref{exactb})
that universal soft masses at $M_X$ imply
$\bxhat^{(1)}(M_X)=\bxhat^{(2)}(M_X)=0$,  using 
Eq.~(\ref{gravnat}) and the fact that it follows 
immediately from Eq.~(\ref{beone:2}) using gauge 
invariance and anomaly cancellation (Eq.~(\ref{anom})) that 
\be
\Tr[\Ycal\gamma^{(1)}]=0.
\ee
Moreover, it is easy to show, using the result for $\beta_{m^2}^{(1)}$ 
from Ref.\cite{jj}, that if 
we work 
consistently at one loop, 
then $\Tr[\Ycal m^2]$ is scale invariant. So if initially 
$\xi = \Tr[\Ycal m^2] = 0$, then $\xi$ remains  zero  
under (one loop) RG evolution. 
With typical universal conditions at $M_X$ with soft masses $m_0$ and 
$M\sim m_0$, $A\sim m_0$,
we find (using 
three loops for $\beta_{\xi}$ and two loops for the other $\beta$-functions) 
that  $\xi \approx 0.001 m_0^2$ at  $M_Z$.

Another favoured set of boundary conditions is those derived from
anomaly mediated symmetry breaking (AMSB)\cite{amsbrefs}. 
Here the soft masses are given by
\be
(m^2)^i{}_j = \frac{1}{2}|m_{\frac{3}{2}}|^2\mu\frac{d\gamma^i{}_j}{d\mu},
\ee
where $m_{\frac{3}{2}}$ is the gravitino mass. In fact, since the AMSB result is
RG invariant, it applies at all scales between $M_X$ and $M_Z$.
We then find from Eqs.~(\ref{exacto}), (\ref{exactb}) 
that up to two loops, we may write
\be
\lf\bxhat=g_1|m_{\frac{3}{2}}|^2\mu{d\over{d\mu}}\Tr[\Ycal (\gamma-\gamma^2 )].
\ee  
Gauge invariance and anomaly cancellation combined with 
Eqs.~(\ref{beone:2}) and (\ref{betwo:2}) yield\cite{xione}
\be
\Tr[\Ycal \gamma^{(1)}] = \Tr[\Ycal(\gamma^{(2)}-(\gamma^{(1)})^2 )]=0,
\ee
and so $\bxhat$ vanishes through two loops. Therefore 
to a good approximation $\xi(M_Z)$ will be given by Eq.~(\ref{xiapp}), 
and once again will be negligible at $M_Z$ if it is zero at $M_X$.

However, if non-universal scalar masses at $M_X$ are contemplated, then the
effects of $\bxhat$ might be significant--as was noted in Ref.~\cite{falk},
for instance. Another context where $\bxhat$ might play a role is that
of non-standard soft-supersymmetry breaking\cite{jja}.
This is because with the non-standard terms (for example
$\phi^2\phi^*$ terms) the result that $\Tr[\Ycal m^2]$ is one-loop
scale invariant
is not preserved. It follows that even with universal boundary
conditions for $m^2$ and $\xi = 0$ at
$M_X$,  $\xi$ becomes non-zero at $M_Z$ even with one-loop running.   
In the current context of the MSSM with the 3rd generation dominating, the
additional soft terms are given by
\be  L^{\rm new}_{\rm soft} = m_{\psi} \psi_{H_1}\psi_{H_2} +
\Atbar\lambda_t H_1^* Q ^c  
+ \Abbar\lambda_b H_2^* Q b^c
+ \Ataubar\lambda_{\tau}H_2^* L \tau^c   + {\rm h.c. }
\label{smssmb}
\ee
Now in Ref.~\cite{jja} we assumed, in fact, that $\xi$ was zero at $M_Z$;
here we explore the
more natural assumption that $\xi = 0$ at the unification scale.      
We follow Ref.~\cite{jja} in dropping the explicit $\mu$-term from the
superpotential, since it can be subsumed into $L^{\rm new}_{\rm soft}$.
With given values at $M_X$ for $m_{\psi}$
and for the universal parameters $A$, $M$ and $m_0$,
and for a given $\tan\beta$, 
we adjust $\Atbar=\Abbar=\Ataubar=\Abar$ (at $M_X$) to
obtain an acceptable electroweak vacuum. As in Ref.~\cite{jja},
we have  made allowance
for radiative corrections by using the tree Higgs minimisation
conditions,  but evaluated at the scale $M_{\rm SUSY}\approx m_0$.
In Fig.~1 we show (for illustrative values of
$M$, $m_{\psi}$ and $A$) the region of the $m_0$, $\tan\beta$
plane where this can be achieved.
\smallskip
\epsfysize= 4in
\centerline{\epsfbox{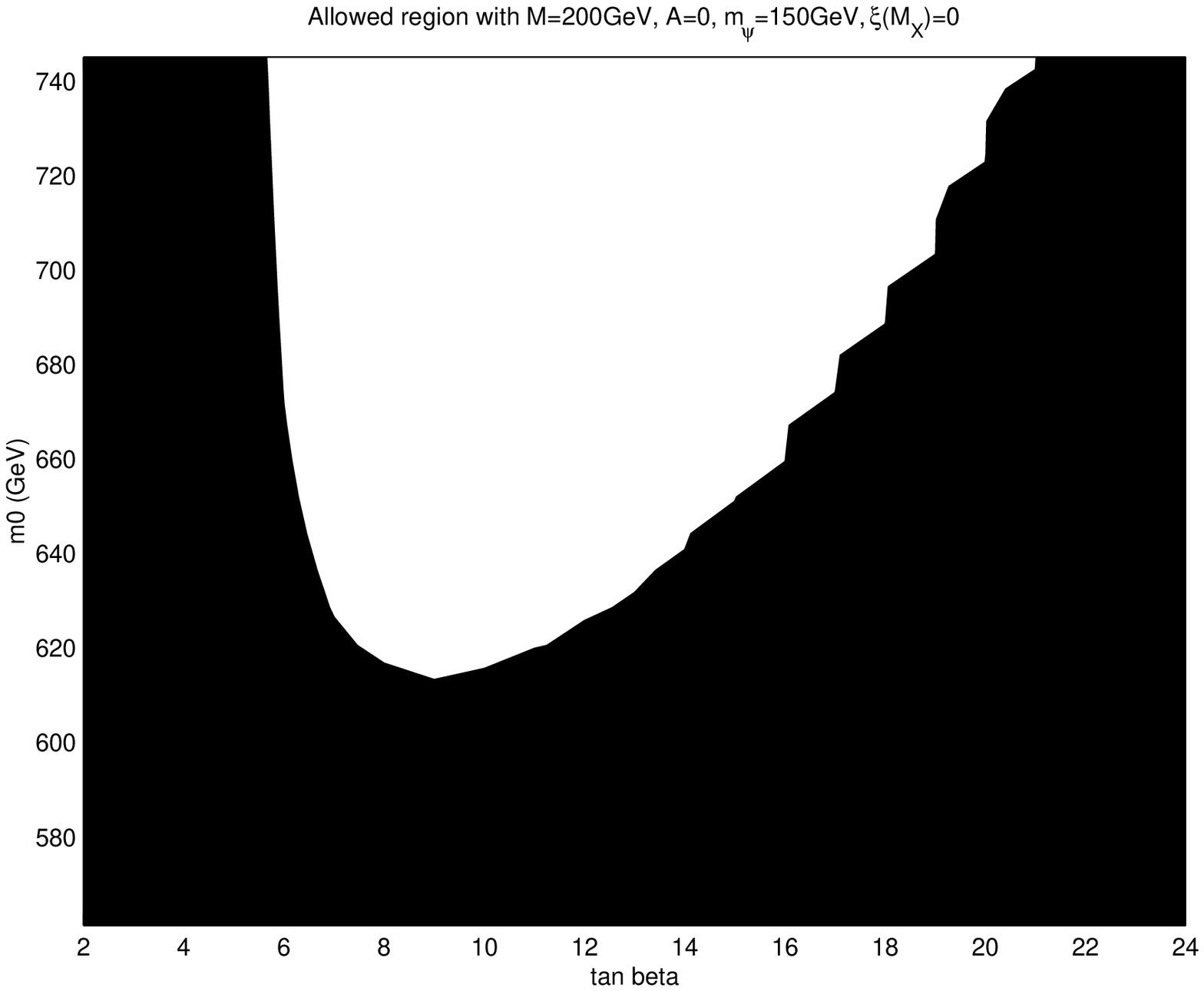}}
\in
{\it \noindent Fig.1:
The region of the $m_0, \tan\beta$ plane corresponding to an acceptable
electroweak vacuum, for $M=200\GeV$, $m_{\psi} = 150\GeV$, $A = 0$ and 
$\xi(M_X)=0$. The shaded
region corresponds to one or more sparticle or Higgs
masses in violation of current
experimental bounds.}
\medskip
\out

For comparison, we show in Fig.~2 the corresponding region for 
$\xi(M_{\rm SUSY})=0$. We notice that it is qualitatively similar, though
slightly larger.
\smallskip
\epsfysize= 4in
\centerline{\epsfbox{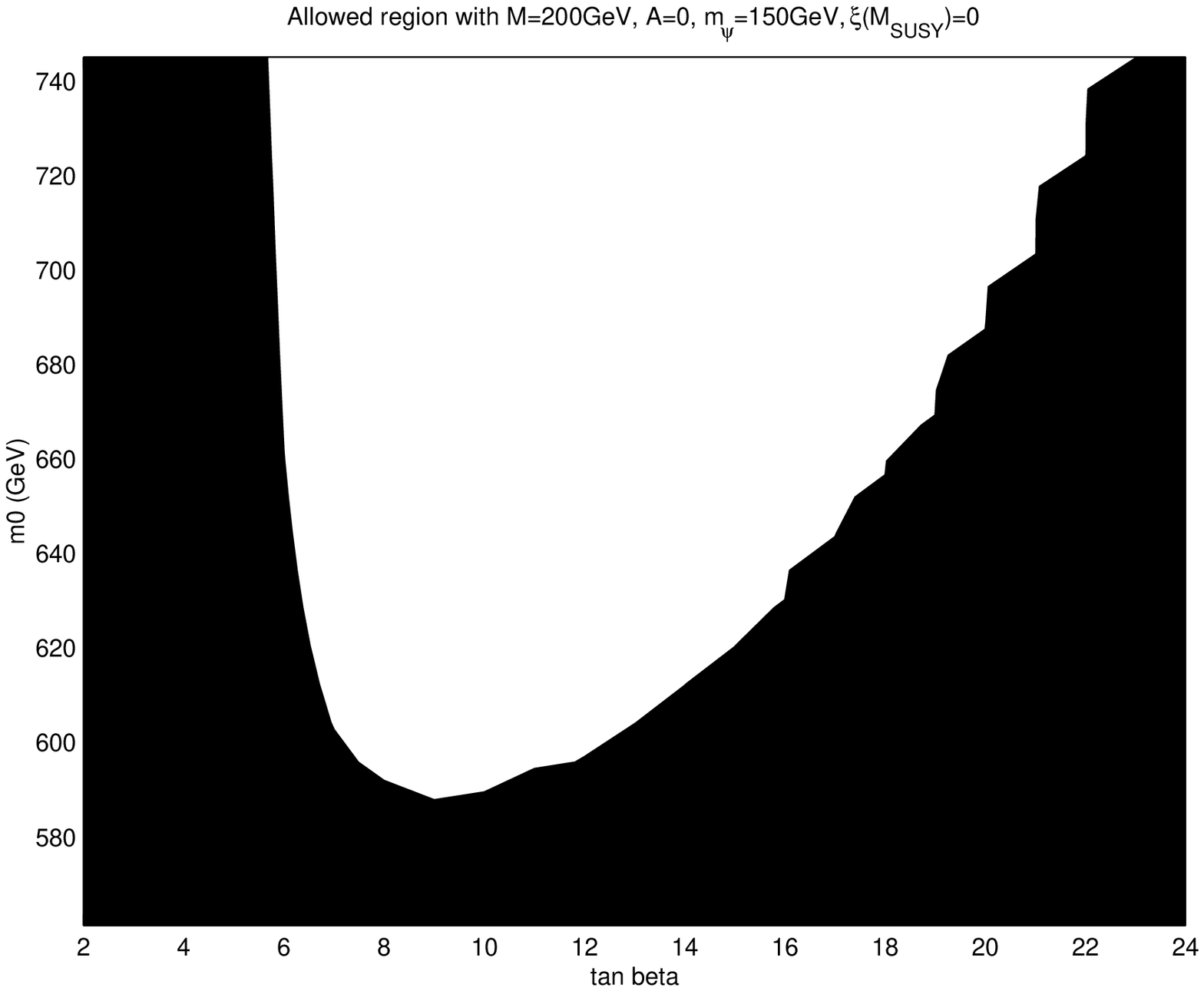}}
\in
{\it \noindent Fig.2:
The region of the $m_0, \tan\beta$ plane corresponding to an acceptable
electroweak vacuum, for $M=200\GeV$, $m_{\psi} = 150\GeV$, $A = 0$ and 
$\xi(M_{\rm SUSY})=0$. The shaded
region corresponds to one or more sparticle or Higgs
masses in violation of current
experimental bounds.}
\medskip
\out

Note that this Figure differs slightly from Fig.~1 of Ref.~\cite{jja}. This is 
because we have incorporated one-loop corrections to the Higgs mass and because 
we have taken account of the increasingly stringent experimental bounds 
(in particular increasing $m_{\psi}$ at $M_X$ to get 
acceptable chargino masses). 
For $m_0=640\GeV$
and $\tan\beta=8$, we find $\Abar=1.07(1.01)\TeV$, $\Atbar(M_{\rm SUSY})\approx
661(627)\GeV$,
$\Ataubar(M_{\rm SUSY})\approx664(630)\GeV$, $\Abbar(M_{\rm SUSY})
\approx491(469)\GeV$. (The pairs of numbers 
correspond to $\xi(M_X)=0$, $\xi(M_{\rm SUSY})=0$ respectively.) 
The spectra obtained for $\xi(M_X)=0$ and for $\xi(M_{\rm SUSY})=0$ 
are given in 
Table~2. We see that there are significant differences, especially in the
masses of $H$, $A$ and $H^{\pm}$. On the other hand the chargino and 
neutralino masses are unaffected, with a LSP neutralino.  
\vskip3em
\begin{center}
\begin{tabular}{|c|c|c|}\hline   
&$\xi(M_X)=0$&$\xi(M_{\rm SUSY})=0$\\ \hline
$\ttil_1 $&639&649\\ \hline
$\ttil_2 $&319&334\\ \hline
$\btil_1 $&604&615\\ \hline
$\btil_2 $&776&780 \\ \hline
$\tautil_1 $&625&639\\ \hline
$\tautil_2 $&663&658\\ \hline
$\util_L $&792&794\\ \hline
$\util_R $&793&785\\ \hline
$\dtil_L $&796&798\\ \hline
$\dtil_R $&781&785\\ \hline
$\etil_L $&664&657\\ \hline
$\etil_R $&632&646\\ \hline
$\nutil_{\tau}$&657&650\\ \hline
$\nutil_e $&659&652\\ \hline  
$h $&116&116\\ \hline
$H $&231&291\\ \hline
$A $&230&290\\ \hline
$H^{\pm} $&244&302\\ \hline  
$\chitil^{\pm}_1 $&120&120\\ \hline
$\chitil^{\pm}_2 $&201&201\\ \hline
$\chitil_1 $&68&68\\ \hline
$\chitil_2 $&116&116\\ \hline
$\chitil_3 $&167&167\\ \hline
$\chitil_4 $&234&234\\ \hline
${\tilde g} $&521&521\\ \hline 
\end{tabular}
\end{center}  
\in
{\noindent \it Table 1: Spectra (in \GeV) 
for $\xi(M_X)=0$ and for $\xi(M_{\rm SUSY})=0$, 
with $M=200\GeV$, $m_0=640\GeV$, $A=0$, $m_{\psi}=150\GeV$ at $M_X$,
and with $\tan\beta=8$. \hfill}
\out
\bigskip

Finally, in Table~2
we give the values of $\xi(M_{\rm SUSY})$ for some typical points in the allowed
region of Fig.~1. We see indeed that $\xi(M_{\rm SUSY})$ is quite sizeable.
\vskip3em
\begin{center}
\begin{tabular}{|c|c|c|}\hline
 $m_0(\GeV)$      & $\tan\beta$           & $\xi(M_{\rm SUSY})(\GeV)^2$   
\\ \hline  
640&8&$-5.07\times10^4$\\ \hline
700&6&$-5.48\times10^4$\\ \hline
700&8&$-5.02\times10^4$\\ \hline  
700&16&$-5.15\times10^4$\\ \hline
800&6&$-5.61\times10^4$\\ \hline
800&8&$-4.90\times10^4$\\ \hline
800&16&$-4.75\times10^4$\\ \hline
\end{tabular}
\end{center}  
\in
{\noindent \it Table 2: Values for $\xi(M_{\rm SUSY})$ with $\xi(M_X)=0$
and with $M=200\GeV$, $A=0$ and $m_{\psi}=150\GeV$ at $M_X$. \hfill}
\out

We have verified that the same results are obtained if we either
(1) Perform the RG evolution in the $\xi$-uneliminated theory and
then eliminate $\xi$ (via its equation of motion) at low energies or
(2) Eliminate $\xi$ at $M_X$, and evolve to low energies with
the (modified) $\xi$-eliminated $\beta$-functions. For a general discussion
of the equivalence of these procedures, see Refs.~\cite{xione}, \cite{xitwo}.

\section{Conclusions}

In this paper we have extended the results of Ref.~\cite{xitwo} for the
renormalisation  of the Fayet-Iliopoulos $D$-term to the  case of a
direct product gauge group, and applied the result to the  MSSM. With
standard soft supersymmetry breaking and universal boundary  conditions
at $M_X$,  then $\xi$ is negligible at low energies  if $\xi(M_X) = 0$.
However with non-standard soft breakings (and/or  non-universal boundary   
conditions for the standard ones) we find  significant effects even for
$\xi(M_X) = 0$. In the  non-standard breaking case, the effect is
especially marked for the masses of $H$, $A$ and $H^{\pm}$, which decrease
significantly when $\xi$  is taken into account.

\acknowledgements
 
Part of this work was done during visits by one of us (DRTJ) 
to SLAC and to the Aspen Center for Physics, and 
was supported in part by SLAC and PPARC and by a Research Fellowship from the
Leverhulme Trust. We thank Steve Martin for encouragement. 
\appendix
\section{general result for several abelian factors}
\label{sec:gen}
In this Appendix we give the general results for the case of a direct product
group with several abelian factors. As we mentioned earlier, the situation is 
complicated by the possibility of ``kinetic mixing''\cite{kin}
between the different
abelian factors. We can accommodate this possibility by introducing a matrix of
couplings for the abelian factors. Suppose that the gauge group is 
 $(U_1)^a\prod_{\alpha=a+1}^nG_{\alpha}$, where the $G_{\alpha}$, 
$\alpha=a+1\ldots n$ are non-abelian. The gauge couplings are then 
$g_{\alpha\beta}$, where $g_{\alpha\beta}=g_{\alpha}\delta_{\alpha\beta}$,
$\alpha=a+1\ldots n$, with a similar form for the gauge $\beta$-functions.
The gaugino masses also form a matrix 
$M_{\alpha\beta}$ with an analogous structure, as do their 
$\beta$-functions.
Suppose the hypercharges of the abelian factors for a
given representation are
$\Ycal_{\alpha}$, $\alpha=1\ldots a$. Then we define 
\be
\Ybar_{\alpha}=\sum_{\beta=1}^a\Ycal_{\beta}g_{\beta\alpha}, 
\quad \alpha=1\ldots a
\ee
and a generalised quadratic Casimir
\be
\Cbar(R)=\sum^a_{\alpha=1}\Ybar_{\alpha}\Ybar_{\alpha}+\sum^n_{\alpha=a+1}
g^2_{\alpha}C(R_{\alpha}).
\ee
The Fayet-Iliopoulos couplings now form a vector $\xi_{\alpha}$,
$\alpha=1\ldots a$, and we have the matrix equation 
\be
\beta_{\xi}=g^{-1}\beta^{g}\xi+\bxhat.
\ee
We can now give the explicit 
general results, starting with the gauge $\beta$-functions 
and anomalous dimension.  
At one loop,
\be
\lf\beta^{g(1)}=g\Qbar 
\ee
where 
\bea
\Qbar_{\alpha\beta}&=&\Tr\left[\Ybar_{\alpha}\Ybar_{\beta}\right],
\quad \alpha,\beta=1\ldots a\nn
\Qbar_{\alpha\beta}&=&g_{\alpha}^2Q_{\alpha}\delta_{\alpha\beta}, 
\quad \alpha=a+1\ldots n
\eea
and
\be
\lf\left(\gamma^{(1)}\right)^i{}_j=P^i{}_j\equiv
\frak12Y^{ikl}Y_{jkl}-2\Cbar(R)^i{}_j.
\ee
At two loops,
\be
\llf\left(\gamma^{(2)}\right)^i{}_j=-\left[Y_{jmn}Y^{mpi}
+2\Cbar(R)^p{}_j\delta^i{}_n
\right]P^n{}_p+2\left(\Qbar_{\alpha\beta}\Ybar_{\alpha}\Ybar_{\beta}
+g_{\alpha}^4Q_{\alpha}C(R_{\alpha})\right)^i{}_j,
\ee
and
\bea
\llf\left(\beta^{g(2)}\right)_{\alpha\beta}&=&-2g_{\alpha\gamma}\Tr
\left[P\Ybar_{\gamma}\Ybar_{\beta}\right],\quad \alpha,\beta=
1\ldots a\nn
\llf\left(\beta^{g(2)}\right)_{\alpha}&=&
2g_{\alpha}^5C(G_{\alpha})Q_{\alpha}
-2g_{\alpha}^3r_{\alpha}^{-1}\Tr\left[P C(R_{\alpha})\right],\quad 
\alpha=a+1\ldots n.
\eea
At three loops we have
\bea
\lllf\left(\beta^{g\DRED(3)}\right)_{\alpha\beta}&=&
g_{\alpha\gamma}
\Bigl\{3Y^{ikm}Y_{jkn}P^n{}_m\left(\Ybar_{\gamma}\Ybar_{\beta}\right)^j{}_i
+6\Tr[P\Ybar_{\gamma}\Ybar_{\beta}\Cbar(R)]\nn
&+&\Tr[P^2\Ybar_{\gamma}\Ybar_{\beta}]
-6\sum_{\kappa,\lambda=1}^a
\Qbar_{\kappa\lambda}
\Tr[\Ybar_{\gamma}\Ybar_{\beta}\Ybar_{\kappa}\Ybar_{\lambda}]
\nn
&-&6\sum_{\kappa=a+1}^ng_{\kappa}^4Q_{\kappa}
\Tr[\Ybar_{\gamma}\Ybar_{\beta}C(R_{\kappa})]\Bigr\}
,\quad \alpha,\beta=1\ldots a\nn
\left(\beta^{g\DRED(3)}\right)_{\alpha}&=&
3r_{\alpha}^{-1}g_{\alpha}^3Y^{ikm}Y_{jkn}P^n{}_mC(R_{\alpha})^j{}_i
+6r_{\alpha}^{-1}g_{\alpha}^3
\Tr\left[PC(R_{\alpha})\Cbar(R)\right]\nn
&+&r_{\alpha}^{-1}g_{\alpha}^3\Tr\left[P^2C(R_{\alpha})\right]
-6r_{\alpha}^{-1}g^3_{\alpha}\sum_{\kappa,\lambda=1}^a
\Qbar_{\kappa\lambda}\Tr[C(R_{\alpha})\Ybar_{\kappa}\Ybar_{\lambda}]
\nn
&-&6r_{\alpha}^{-1}g^3_{\alpha}
\sum_{\kappa=a+1}^ng_{\kappa}^4Q_{\kappa}\Tr[C(R_{\alpha})C(R_{\kappa})]
- 4r_{\alpha}^{-1}g_{\alpha}^5C(G_{\alpha})\Tr\left[PC(R_{\alpha})\right]\nn
&+&g_{\alpha}^7Q_{\alpha}C(G_{\alpha})\left[4C(G_{\alpha})-Q_{\alpha}\right],
\quad \alpha=a+1\ldots n. 
\eea
For the Fayet-Iliopoulos couplings we have at one loop
\be
\lf\left[\bxhat^{(1)}\right]_{\alpha}=\Tr\left[\Ybar_{\alpha}m^2\right],
\quad \alpha=1\ldots a
\ee
and at two loops
\be
\lf\left[\bxhat^{(2)}\right]_{\alpha}=-4\Tr\left[\Ybar_{\alpha}m^2\gamma^{(1)}
\right].
\ee
Finally, 
\bea
\lllf\left(\bxhat^{(3)\DREDp}\right)_{\alpha} &=& 
-6(\lf)^2\Tr\left[\Ybar_{\alpha} m^2
\gamma^{(2)}\right]-4\Tr\left[\Ybar_{\alpha}WP\right]-
\frac{5}{2}\Tr\left[\Ybar_{\alpha}HH^{\dagger}\right]\nn
&+& 2\Tr\left[\Ybar_{\alpha}P^2m^2\right]
-24\zeta(3)\Tr\left[\Ybar_{\alpha} W\Cbar(R)
\right]\nn
&+& 12\zeta(3)\Tr\left[\Ybar_{\alpha}H
\Cbar^{M*}(R)
+ \hbox{c.c.}\right]-96\zeta(3)
\Tr[\Ybar_{\alpha} \Cbar^{MM^*}(R)\Cbar(R)]\nn
&-&24\zeta(3)\left\{\Tr[\Ybar_{\alpha} \Cbar^{M}(R)\Cbar^{M*}(R)]+\hbox{c.c.}
\right\},
\eea
where
\bea
W^i{}_j &=&
(\frac{1}{2}Y^2m^2 +\frac{1}{2}m^2Y^2 +h^2)^i{}_j
+2Y^{ipq}Y_{jpr}(m^2)^r{}_q -8\Cbar^{MM^*}(R)^i{}_j,\nn
H^i{}_j&=&h^{ikl}Y_{jkl}+4\Cbar^M(R)^i{}_j,\nn
\Cbar^M(R)&=&\sum_{\alpha,\beta=1}^aM_{\alpha\beta}\Ybar_{\alpha}\Ybar_{\beta}
+\sum_{\alpha=a+1}^ng_{\alpha}^2M_{\alpha}C(R_{\alpha}), \nn
\Cbar^{MM^*}(R)&=&\sum_{\alpha,\beta=1}^a(MM^*)_{\alpha\beta}\Ybar_{\alpha}
\Ybar_{\beta}+\sum_{\alpha=a+1}^nM_{\alpha}M^*_{\alpha}g_{\alpha}^2
C(R_{\alpha}).
\eea


\begin{references}
\bibitem{xione} I.~Jack and D.R.T.~Jones, \plb473 (2000) 102
\bibitem{xitwo} I.~Jack, D.R.T.~Jones and S.~Parsons, \prd62 (2000) 125022
\bibitem{gouv} A. de Gouv\^ea,
A. Friedland and H. Murayama, \prd 59 (1999) 095008  
\bibitem{kin} B.~Holdom, \plb166 (1986) 196\semi
F.~del Aguila, G.D.~Coughlan and M.~Quir\'os, \npb307 (1988) 633\semi
F.~del Aguila, Acta Phys. Polon. {\bf B}25 (1994) 1317\semi
F.~del Aguila, M.~Cveti\v c and P.~Langacker, \prd52 (1995) 37\semi
K.S.~Babu, C.~Kolda and J.~March-Russell, \prd54 (1996) 4635\semi
K.R.~Dienes, C.~Kolda and J.~March-Russell, \npb492 (1997) 104\semi
K.S.~Babu, C.~Kolda and J.~March-Russell, \prd57 (1998) 6788
\bibitem{jjn} I.~Jack, D.R.T.~Jones and C.G.~North, \npb486 (1997) 479
\bibitem{jj} I.~Jack and D.R.T.~Jones, \plb 333 (1994)
372
\bibitem{jjmvy}I.~Jack, D.R.T.~Jones,
S.P.~Martin, M.T.~Vaughn and Y.~Yamada, \prd50 (1994) R5481
\bibitem{bjork}J.E.~Bj\"orkman and D.R.T.~Jones, \npb 259 (1985) 533
\bibitem{amsbrefs}L. Randall and R. Sundrum,  \npb 557 (1999) 79\semi
G.F. Giudice, M.A. Luty, H. Murayama and  R. Rattazzi,
JHEP 9812 (1998) 27\semi
A. Pomarol and  R. Rattazzi, JHEP 9905 (1999) 013\semi
T. Gherghetta, G.F. Giudice and  J.D. Wells, \npb 559 (1999) 27\semi
M.A. Luty and R. Rattazzi, JHEP 9911 (1999) 001\semi
Z. Chacko, M.A. Luty, I. Maksymyk and E. Ponton, JHEP 0004 (2000) 001\semi
E. Katz, Y. Shadmi and Y. Shirman, JHEP 9908 (1999) 015\semi  
I. Jack and D.R.T.~Jones, \plb 465 (1999) 148\semi
J.L.~Feng and T.~Moroi, \prd 61 (2000) 095004\semi%
G.D.~Kribs, \prd 62 (2000) 015008\semi%
S.~Su, \npb 573 (2000) 87\semi%
J.A.~Bagger, T. Moroi and E. Poppitz, JHEP 0004 (2000) 009\semi
R. Rattazzi et al, \npb576 (2000) 3\semi%
F.E.~Paige and J. Wells, hep-ph/0001249%
\bibitem{falk} T. Falk, \plb456 (1999) 171
\bibitem{jja}I. Jack and D.R.T.~Jones, \plb457 (1999) 101; 
\prd 61 (2000) 095002
\end{references}
\end{document}